\documentclass[fleqn,twoside]{article}
\usepackage{espcrc2}
\usepackage{amsmath,amssymb,bm}

\usepackage{graphicx}
\usepackage[figuresright]{rotating}

\title{
On the lattice construction of electroweak gauge theory}

\author{Yoshio Kikukawa\address[NGY]{Department of Physics, Nagoya
 University, Nagoya 464-8602, Japan},
        Yoichi Nakayama\addressmark ,
        Hiroshi Suzuki\address{Department of Mathematical Sciences,
        Ibaraki University, Mito 310-8512, Japan}
}
\begin{document}

\begin{abstract}
Based on the Ginsparg-Wilson relation, a gauge invariant formulation 
of electroweak SU(2)$\times$U(1) gauge theory on the lattice is considered.
If the hypercharge gauge coupling is turned off in the 
vacuum sector of the U(1) gauge fields, 
the theory consists of four left-handed SU(2) 
doublets and it is possible, as in vector-like theories, 
to make the fermion measure defined globally 
in all topological sectors of SU(2).
We then try to incorporate U(1) gauge field, following
L\"uscher's reconstruction theorem.
The global integrability condition is proved
for ``gauge loops'' in the space of the %admissible 
U(1) gauge fields
with arbitrary SU(2) gauge field fixed in the background.
For ``non-gauge loops'', however, 
the proof is given 
so far 
only for the classical SU(2) instanton backgrounds.
\vspace{1pc}
\end{abstract}

\maketitle

\section{An approach to electroweak theory}

Recently, gauge-covariant and local lattice Dirac operators satisfying
the Ginsparg-Wilson relation\cite{Ginsparg-Wilson}
has been
constructed\cite{PH-fixed-point-D,HN-chiralsym,PH-KJ-ML-locality},
and this opens the possibility to formulate chiral gauge theories 
on the lattice with exact gauge invariance.
Indeed, L\"uscher has given a constructive proof of the existence
of Weyl fermion measure in anomaly-free abelian 
chiral gauge theories \cite{ML-abelian}.
The same author has also examined the construction 
of generic non-abelian chiral gauge theories and has formulated 
the reconstruction theorem for the fermion measure \cite{ML-nonabelian}.
The purpose of this article is to examine the possible extension
of the construction to 
electroweak SU(2)$\times$U(1) gauge theory.

Electroweak theory is the chiral gauge theory 
of left-handed leptons and quarks in SU(2) doublet
and right-handed quarks in SU(2) singlet. Taking into account
of the color degrees of freedom, there are four doublets 
in each generation.
In order to construct this theory on the lattice,
we 
adopt the definition of lattice Weyl fermion based on
Ginsparg-Wilson Dirac operator
\cite{ML-chiralsym,RN-HN-overlap,FN-lat98,ML-abelian}:
\begin{equation}
 \gamma_5 D + D \hat \gamma_5 = 0, \qquad 
\hat\gamma_5 \equiv \gamma_5(1-aD),
% \gamma_5 D + D \gamma_5 = a D \gamma_5 D,
\end{equation}
\begin{equation}
 \psi_L(x) = \hat P_- \psi_L(x) , \quad \hat P_- = (1-\hat\gamma_5) / 2 .
\end{equation}
The projection for $\bar\psi$ is defined as usual with $\gamma_5$. 
Gauge fields are given by link variables
$U(x,\mu) = U^{(2)}(x,\mu) \otimes  U^{(1)}(x,\mu)$.
We assume the lattice to be finite with periodic boundary conditions
in all directions.
We require that U(1) and SU(2) components of link variables
satisfy the so-called admissibility conditions respectively,
which ensure the locality and smoothness of the Dirac operator
\cite{PH-KJ-ML-locality}.
%:
%\begin{equation}
%\parallel 1 - R [ U^{(n)}_{\mu\nu}(x) ] \parallel <  \epsilon \qquad 
%(n = 2,1),
%\end{equation}
%where 
%$U^{(n)}_{\mu\nu}(x)$ is the plaquette variable
%and $R$ denotes a representation of the gauge group. 
%$\epsilon$ is some sufficiently small
%constant\cite{ML-topology}.
%%For the U(1) gauge field we adopt the following condition, 
%%\begin{equation}
%%|6 Y| \cdot | F_{\mu\nu}(x)|<  \frac{\pi}{3}, \
%%F_{\mu\nu}(x)= \frac{1}{i}\ln \{U^{(1)}_{\mu\nu}(x)\}.
%%\end{equation}
Then the space of the admissible SU(2)$\times$U(1) gauge
fields is divided into the topological sectors\cite{ML-topology}, 
each one is the product of a U(1) topological sector 
and an SU(2) topological sector.

Our approach to electroweak theory on the lattice
relies on the specific feature of the original theory.
In the vacuum sector of the U(1) gauge fields
where 
%no non-trivial magnetic flux exists and 
any configuration can be deformed smoothly to the trivial one 
$U(x,\mu)=1$,
we can turn off the hypercharge gauge coupling. Then
the theory can be regarded as 
vector-like due to the pseudo reality of SU(2).
%a pair of left-handed doublets % Weyl fermions %in $\underbar{2}$ 
%is equivalent 
%to a single Dirac fermion in the same representation
%through charge conjugation.
It is indeed possible
%as in vector-like theories, 
to make the fermion measure 
%independent of the choice of the basis and thus 
defined globally
in all topological sectors of 
SU(2) by the following choice of the basis for a pair of 
doublets $(a,b)$\cite{ML-abelian,HS-real}:
\begin{eqnarray}
\label{eq:su2-basis-global-a}
w_j^{(a)}(x) &=& u_j(x),   \\
\label{eq:su2-basis-global-b}
w_j^{(b)}(x) &=& \left( \gamma_5 C^{-1} \otimes 
i \sigma_2 \right) \, \left[ u_j(x) \right]^\ast ,
\end{eqnarray}
where $\hat P_- u_j(x) = u_j(x)$. This fact 
implies the cancellation of Witten's SU(2) anomaly 
(cf. \cite{OB-IC}).
%One may also adopt the 
%so-called symplectic basis for real representations 
%considered by one of the authors \cite{}.
Given the basis for the SU(2) doublets defined globally, 
one may try to extend the fermion measure 
to incorporate the U(1) gauge field
following the reconstruction theorem \cite{ML-nonabelian},
as far as we concern the vacuum sector of 
the U(1) gauge fields. 

%In this reason,  
%we will concentrate 
%to the vacuum sector of the U(1) gauge fields
%in the following discussion.

\section{A choice of the fermion measure}

According to the reconstruction theorem, 
the first step to obtain the Weyl fermion measure
is to construct 
the measure term ($\eta_\mu \equiv \delta U_\mu {U_\mu}^{-1}$): 
\begin{equation}
\mathfrak{L}  = a^4 \sum_x \left\{
\eta^{(2)}_\mu(x)  j^{(2)}_\mu(x)  + \eta^{(1)}_\mu(x) j^{(1)}_\mu(x)
\right\} ,
\end{equation}
where $j^{(n)}_\mu(n=2,1)$ should satisfy the 
anomalous conservation law and the integrability condition,
and should be defined smoothly over the space of the gauge fields.
This question can be mapped to the equivalent 
local cohomology problem in the 4+2 dimensions.
Two of the authors have examined the cohomology problem
for electroweak theory and have shown that the 
currents $j^{(n)}_\mu(n=2,1)$ with the desired properties
can be constructed in the infinite volume\cite{YK-YN-anomaly}.
Then the remaining issue is to obtain the 
currents $j^{(n)}_\mu(n=2,1)$  in the finite volume.
%two of the authors have performed 
%the cohomological analysis of gauge anomaly\cite{Kikukawa:2001kd},
%following the $4+2$-dimensional approach of \cite{Luscher:2000un}.
%Thanks to the pseudo reality of SU(2) (implying the absence of 
%the SU(2) non-abelian gauge anomaly) and 
%the assignment of the hypercharges in the theory, 
%it is indeed possible to establish the exact cancellation of the 
%gauge anomaly of the mixed type SU(2)$^2\times$U(1)
%at finite lattice spacing $a$, 
%as well as the cubic 
%one U(1)$^3$ \cite{}.
The similar problem has already been solved by L\"uscher in the
case of abelian chiral gauge theories\cite{ML-abelian},
and that procedure can actually be applied 
to the U(1) part of the current $j_\mu^{(1)}$ in our case,
if we regard the SU(2) gauge field as a background.
%
%Where we start with the fact of gauge anomaly cancellation on infinite
%volume lattice in the same context\cite{YK-YN-anomaly}, where the essential is
%to use U(1) hypercharge assignments and pseudo reality of SU(2).
%$j_\mu^{(1)}$ %, the U(1) part of the current, 
%so obtained satisfies the required conditions
%with respect to the U(1) gauge field.
The resulted current is local and smooth with respect to the SU(2) 
gauge field and is invariant under the SU(2) gauge transformation.
%Thus the first problem can be partially solved. 

% For notations and terminologies, see
% \cite{ML-abelian,ML-nonabelian}.

Given the smooth current $j_\mu^{(1)}(x)$ 
and the ``global'' measure for $U^{(2)}\otimes 1$ 
(the associated Weyl fermion basis $\left\{w_j(x)\right\}$),
we may consider a choice of the fermion measure as follows:
let us consider a smooth curve in the space of the gauge fields,
along which the U(1) gauge field is interpolated from the 
trivial configuration $U^{(1)}_0= 1$ to a certain non-trivial 
configuration $U^{(1)}_1 = U^{(1)}$ in the vacuum sector,
while the SU(2) gauge field is fixed to an arbitrary 
configuration $U^{(2)}$ in a given topological sector.
Along the curve $\{U^{(2)}\otimes U^{(1)}_t\}_{0\leq t\leq 1}$,
we introduce the Wilson line $W^{(1)}$ using $j_\mu^{(1)}$, 
%\begin{eqnarray}
%W=\exp\biggl(i\int_0^1{\rm d}t\, \sum_x 
%\eta_\mu^{(1)} j_\mu^{(1)}\biggr),\\
%\quad a\eta_\mu^{(1)}(x)=\partial_tU_t^{(1)}(x,\mu)U_t^{(1)}(x,\mu)^{-1},
%\end{eqnarray}
and the evolution operator $Q_t^{(1)}$ ($0\leq t\leq 1$).
%\begin{equation}
% P_t = Q_t P_0 Q_t^{-1} ,\quad P_t = \hat P_{-} |_{U^{(2)}\otimes U_t^{(1)}}
%\end{equation}
Then we define a basis for $U^{(2)}\otimes U^{(1)}_1$ as follows:
\begin{eqnarray}
\label{eq:the-choice-of-basis}
 v_j  \,  =
\begin{cases}
 Q_1^{(1)} w_1 {W^{(1)}}^{-1}  &\mbox{if}\quad j=1 \\
 Q_1^{(1)} w_j           &\mbox{otherwise} . 
\end{cases}
\end{eqnarray}
%For the reference configuration $U^{(2)} \otimes  1$,  
%we take the basis $\left\{w_j\right\}$ globally constructed as above.

We can check directly 
the local properties of the fermion measure 
obtained from the basis (\ref{eq:the-choice-of-basis}),
by evaluating the measure term in the neighbor of $U^{(2)}\otimes U^{(1)}$.
We can see that 
both currents $j^{(2)}_\mu$ and $j^{(1)}_\mu$
are obtained in this case
and they indeed satisfy the integrability condition and the
anomalous conservation law. Thus this choice of the basis gives
a consistent fermion measure at least locally. 

\section{Global integrability for U(1) loops}

For the fermion measure defined with
the basis (\ref{eq:the-choice-of-basis}) 
to be consistent globally, 
it is required that the Wilson line $W^{(1)}$ should satisfy
the global integrability condition 
\begin{equation}
  W^{(1)} = \det (1-P_0 + P_0 Q_1^{(1)})
  \label{eq:global-integrability}
\end{equation}
($P_t = \hat P_{-} |_{U^{(2)}\otimes U_t^{(1)}}$)
for all closed loops in the space of the U(1) gauge fields
with arbitrary SU(2) gauge field fixed in the background.
We next examine this global integrability condition in this section.

The admissible U(1) gauge fields in the vacuum sector can 
be parametrized as 
%U(1) configurations connected to U(1) vacuum
%can be classified as follows:
%\begin{equation}
% U(x,\mu)= \dot{U}(x,\mu)\exp\{ i A^T_\mu(x)\}
%\end{equation}
%where $A^T$ denotes contractible component of the field and
%$\dot{U}$ is the configuration with anywhere vanishing field tensor
%and such configuration can be written as
%is product of U(1) phases and can be described as
\begin{align}
 U^{(1)}(x,\mu) &= \Lambda (x) w(x,\mu)\Lambda (x+\hat\mu)^{-1} ,\\
w(x,\mu) &=
  w_\mu \delta_{\tilde x_\mu 0} ,
\end{align}
up to contractible components, where 
$\Lambda(x) \in {\rm U(1)}$, $w_\mu \in {\rm U(1)}$.
There are two types of non-trivial loops in the 
space of the U(1) gauge fields:
the first one is related to the gauge degrees of freedom
$\Lambda(x)$ and referred as ``gauge loops'',
and the second one is ``non-gauge loops'' related to 
%the Wilson lines
$w_\mu$.
Then we should examine the global integrability condition for 
such non-trivial loops while the SU(2) gauge field is kept fixed.

For the gauge loops
\begin{align}
U^{(1)}_t(x,\mu) &= \Lambda_t(x)\Lambda_t(x+\hat\mu)^{-1},\\
\Lambda_t(x) &= \exp \{2\pi i t \delta_{\tilde{x}\tilde{y}}\},
\end{align}
we can prove the global integrability 
in the similar manner as 
abelian chiral gauge theories:
we set $\eta^{(1)}_\mu(x) = -\partial_\mu \delta_{\tilde{x}\tilde{y}}$,
and $\eta^{(2)}_\mu(x)=0$, and then the anomalous conservation law implies
that the Wilson loop is given by 
the U(1) part of the gauge anomaly as 
\begin{equation}
\label{eq:wilson-loop-for-gauge-loops}
 W = \exp \{i 2\pi {\cal A}^{(1)}(y)_{t=0}\}.
\end{equation}
On the other hand, the twist (the determinant on the r.h.s. of
(\ref{eq:global-integrability})) can be 
evaluated with the formula $(t_k = k /n )$
\begin{equation}
  \lim_{n\rightarrow \infty}
\det\left(1-P_{t_0}+P_{t_n}P_{t_n-1}\cdots P_{t_0} \right)
 \label{eq:discrete-def-of-twist}
\end{equation}
and $P_t = R[\Lambda_t] P_0 R[\Lambda_t]^{-1}$
to reproduce the r.h.s. of~(\ref{eq:wilson-loop-for-gauge-loops}).
As we can see, this proof holds 
with any SU(2) gauge field in the background.

For the non-gauge loops
\begin{equation}
U^{(1)}_t (x,\mu) = \exp \{2\pi it \delta_{\mu\nu}\delta_{\tilde{x}_\nu 0}\},
\end{equation}
we could prove the global integrability condition by 
using the reflection property of the current $j_\mu^{(1)}$
in the similar manner as abelian chiral gauge theories:
\begin{equation}
\label{eq:current-reflection}
 j^{(1)}_\mu(x)|_{t\rightarrow 1-t} = -j^{(1)}_\mu (-x+\hat I -\hat\mu),
\end{equation}
where the center of reflection is $\hat I/2$ ($\hat I = (1,1,1,1)$).
With the SU(2) gauge field in the background, however,
we need to require that 
the SU(2) gauge field should have the following reflection symmetry
up to gauge transformation,
\begin{equation}
% U^{(2)}(x,\mu)\rightarrow
 U^{(2)}(-x+\hat I -
 \hat\mu ,\mu)^{-1} \cong U^{(2)}(x,\mu).
 \label{eq:background-has-reflection-symmetry}
\end{equation}
For this restricted class of the SU(2) gauge fields, 
the Wilson line $W^{(1)}$ turns out to be unity,
because of (\ref{eq:current-reflection}).
On the other hand, the twist can be evaluated to be unity,
using the reflection property of the projection operator:
$P_t = {\mit\Gamma} P_{1-t}{\mit\Gamma}^{-1}$ where
${\mit\Gamma}\psi(x) \equiv \gamma_5 \psi(-x)$.

The classical instantons mapped on to a sufficiently large lattice
(centered at $\hat{I}/2$, the middle of the lattice sites) 
indeed possess such reflection invariance, as well as the 
trivial gauge field $U^{(2)}=1$.
It is conceivable that each topological sector of the
SU(2) gauge fields has such a reflection invariant configuration.
So far, the global integrability condition for the non-gauge loops
can be shown only for these restricted SU(2) gauge fields.
%
%Note that this proof for non-gauge loops only stands for reflection
%invariant SU(2) background.

\section{Discussions}

Our approach to electroweak theory on the lattice 
refers to the trivial U(1) configuration $U^{(1)}=1$,
for which the fermion measure can be constructed globally in all 
topological sectors of SU(2),
and therefore is restricted to the vacuum sector 
of the admissible U(1) gauge fields.
It is not clear yet
that such a fermion measure exists also in the U(1) magnetic flux sectors.

Our construction of the measure term is incomplete 
in the sense that 
the SU(2) part $j_\mu^{(2)}$ is not constructed
explicitly and it is not clear if $j_\mu^{(2)}$ could be defined globally.
The latter condition seems to be equivalent to 
the global integrability condition of $j_\mu^{(1)}$ 
for both gauge and non-gauge loops 
with any SU(2) background.
To show these conditions, it seems necessary to 
clarify the topological structure of the space of 
the admissible SU(2) gauge fields and to find
the parametrization of the SU(2) link variables.

%
%
%Once global integrability  for U(1) loops with each arbitrarily fixed
%SU(2) configurations in the sector is proved,
%then uniqueness of the measure is guaranteed and
%we have global measure. We examine integrability condition in this section.
%
%To construct lattice electroweak theory, we proposed
%a way to construct the fermion  measure.
%We have shown our measure has desired local properties.
%And, if global integrability for U(1) direction with all 
%fixed SU(2) background is ensured, then it gives global measure
%in each sectors connected to U(1) vacuum.
%
%We have shown the global integrability for U(1) ``gauge loops''
%with arbitrary SU(2) configuration. For remaining non-trivial loops
%(``non-gauge loops'') we showed the integrability with some special
%reflection invariant SU(2) configurations.
%
%The construction in this way is still in progress, but we stress that
%if the global integrability for non-gauge loops with arbitrary SU(2)
%background is shown, then we can construct electroweak gauge theory on
%the lattice along this line.

\end{document}